\documentclass{elsart}

\usepackage{graphicx}
\usepackage{amssymb}

\begin{document}
\begin{frontmatter}
\title{On the characterisation of  a Bragg spectrometer with X-rays from an ECR source}

\author[ioannina]{D.~F.~Anagnostopoulos},
\author[atomki]{S.~Biri},
\author[juelich]{D.~Gotta},
\author[wien]{A.~Gruber},
\author[paris]{P.~Indelicato},
\author[psi]{B.~Leoni},
\author[wien]{H.~Fuhrmann},
\author[psi]{L.~M.~Simons\corauthref{cor}},
\corauth[cor]{Corresponding author. Address: L.~M.~Simons, WMHA/B24, CH 5232 Villigen, PSI}
\ead{leopold.simons@psi.ch}
\author[psi]{L.~Stingelin},
\author[psi]{A.~Wasser},
\author[wien]{J.~Zmeskal}

\address[ioannina] {Department of Material Science 
and Engineering, University of 
Ioannina, Ioannina, Greece}
\address[atomki]{Institute of Nuclear Research (ATOMKI), Debrecen, 
P.O. Box 51, H4001,Hungary}
\address[juelich]
{Institut f\"{u}r Kernphysik, Forschungszentrum J\"{u}lich, D-52425 J\"{u}lich,
 Germany}
 \address[wien]
{Institut f\"{u}r Mittelenergiephysik, Austrian Acad. of Sci., Vienna}
\address[paris]
{Laboratoire Kastler Brossel, \'Ecole Normale Sup\'erieure et Universit\'e
 P. et M. Curie, Case 74, 4 place Jussieu, 
F-75252, Cedex 05, France}
\address[psi]
{Paul Scherrer Institut, Villigen PSI, CH5232 Villigen, Switzerland}

\begin{abstract}

Narrow  X-ray lines from helium-like argon emitted from  a dedicated 
ECR source have been used to determine the response function of a Bragg
 crystal spectrometer equipped with   large area spherically bent 
silicon (111) or  
 quartz (10$\bar{1}$)   crystals.  The measured spectra are compared 
with simulated ones created by a ray-tracing code based on the  
expected theoretical crystal's rocking curve and the geometry 
of the experimental set-up.

\end{abstract}
\begin{keyword}X-ray spectroscopy, Exotic atoms, ECR source
\PACS 29.30.Kv, 36.10.-k, 29.25.Ni
\end{keyword}
\end{frontmatter}


\section{Introduction}
In an ongoing experiment at the Paul Scherrer Institute (Switzerland) 
 the strong interaction broadening of the ground state level of 
pionic hydrogen of about 1 eV should be determined with an accuracy 
of less than 10 meV  by measuring  the line shape  of
 X-ray transitions feeding the atomic 1s state \cite{prop,biri,caen}. 
The energies of the  pionic hydrogen X-rays in question are 2.436, 
2.886 and   3.043 keV  for the 2p$\to$1s, 3p$\to$1s 
and  4p$\to$1s transitions,
respectively. 
The experiment uses a Bragg spectrometer in Johann mounting \cite{johann}
and was equipped during recent measurements  with 
 spherically bent silicon or quartz crystals having a diameter of 
100 mm and a thickness of 0.3 mm or 0.2 mm, respectively. They were 
mounted by optical contact on glass lenses of ultimate quality. The curvature 
radius $R_C$ of the bent crystals is $R_C$= 2982.5$\pm 0.6$ mm.  
 For these parameters 
 bent crystal theory predicts a negligible influence of 
the bending process on the crystal's rocking curve \cite{forster}. 
Imperfections  in the crystal material itself  and defects produced 
during the fabrication  could lead to deviations which should be 
determined experimentally. In   order to meet the requirements imposed
 by the pionic hydrogen  experiment a  determination of the spectrometer's 
response function requires at least 20000 events in the calibration line.
 Moreover the measurement should be performed under conditions 
 equivalent to the real  experiment.

In earlier experiments a rough energy calibration was done 
by using X-rays from geometrically extended fluorescence targets excited 
 by means of  X-ray tubes \cite{shimoda}. A determination of the response function is out of reach 
with this method as  the  natural line widths are 
much broader than the resolution of the spectrometer.
 In addition the  shape of these lines is distorted by 
satellite transitions.

Alternating the pionic hydrogen measurement in regular 
 intervals with calibration experiments using narrow X-ray lines 
 from other exotic atoms would be ideal. The measured low 
 count rates of about 100/hour, however, render such an approach unfeasible.

The use of X-rays emitted from highly ionized ions  therefore offers 
an appropiate  solution albeit in a remote 
 measurement.  It makes use of the fact that for low to medium Z atoms the  E1 
X-rays from hydrogen-like ions  as well as  M1 X-rays from helium-like 
ions have natural line widths negligibly small compared to the 
expected resolution of the Bragg crystals. To give an example, the lifetime
 of the  M1 transition of helium-like argon used mostly in the present 
study is 0.21 $\mu$s  \cite{paul1}. The required  X-rays are emitted 
in copious quantities from Electron Cyclotron Resonance (ECR) or 
Electron Beam Ion Trap (EBIT) sources. ECR sources are 
 preferable as the X-rays are emitted from a geometrically extended 
 region important for a Johann set-up whereas  EBIT sources 
produce X-rays in the region 
of the very compressed electron beam only, which would require an 
 observation perpendicular to the electron beam cutting down the 
 accepted intensity considerably. Decisive, however, is the low 
 kinetic energy of the ions in ECR sources  on the level of less 
than 1eV only, which had been determined 
at different ECR sources with high resolution optical  
 spectroscopy \cite{bern,sadeghi}. 
 From this a Doppler broadening of less than 
40 meV  can be expected for transitions around 3 keV energy 
 in  helium-like argon.  
In principle EBIT sources could  be tuned to low kinetic energies of the 
 captured ions by reducing the voltage of the drift tube, which,
 however, results in a drastic reduction in intensity \cite{silver}.

The intention of this paper is much different from earlier 
publications using crystal spectrometers in combination with 
sources of highly charged ions. They dealt with  a determination of
 plasma properties  
leading to a deeper understanding of the processes governing the 
physics of  ECR  sources  \cite{douy,gru} or of Tokamak devices \cite{marchuk,bitter} as well as of plasma microsources \cite{faenov}. At an EBIT source 
it became possible recently to determine even the polarization 
 of x-ray emission lines with  bent crystal spectrometry \cite{robbins}.  
 In the present paper, for the first time the characterisation of a crystal spectrometer with a
narrow X-ray line from an almost completely ionized 
atom is described. For this purpose the M1  transition from  helium-like
 argon at 3.104 keV  ($\lambda$=3.994 \AA\cite{paul1} close in energy to the 4p$\to$1s transition
 in pionic hydrogen was used.

\section{Experimental set-up}
\begin{figure}[t]
\centering
\includegraphics[angle=0,scale=0.7]{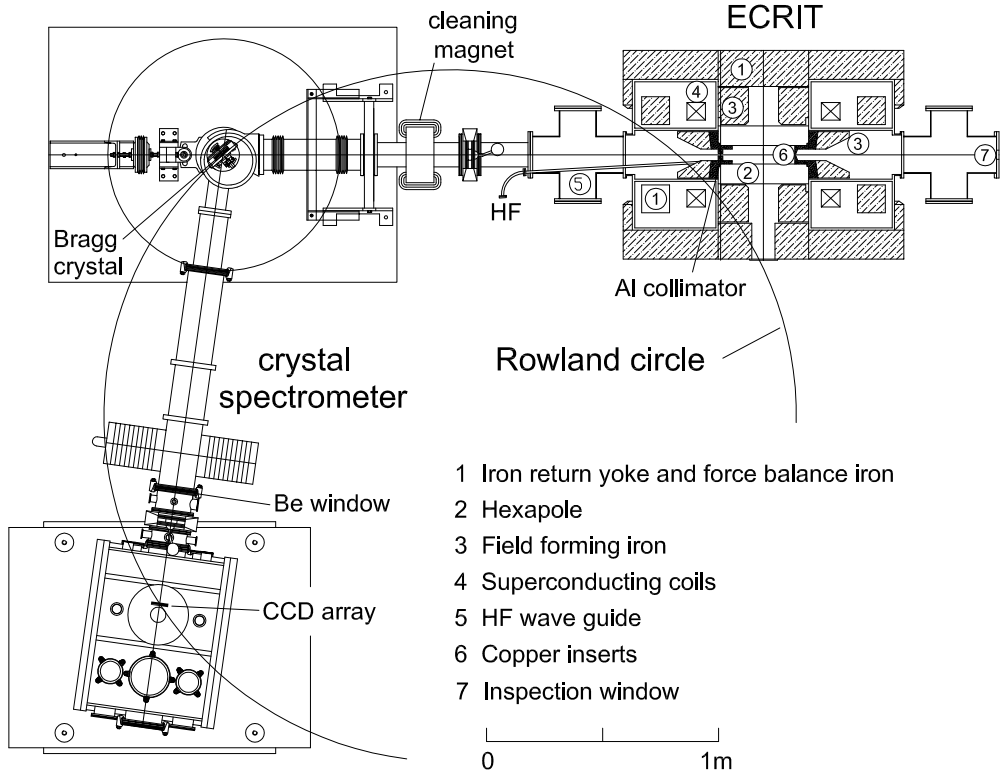}
\caption[]{
Set up of the PSI ECRIT together with the Bragg crystal spectrometer 
}
\label{fig:ecrit+spect}
\end{figure}

The experiment as depicted in Figure \ref{fig:ecrit+spect} can be subdivided
 into three parts:

\begin{itemize}
\item 
 The Electron Cyclotron Resonance Ion Trap  (ECRIT)  
 consists out of a superconducting 
split coil magnet, which together with special iron inserts, provides the 
mirror field configuration, an Advanced ECR source - Updated version (AECRU-U) 
style permanent  hexapole magnet  and a 6.4 GHz 
power regulated microwave emitter \cite{xie}.  The mirror field
 parameters provide 
one of the highest mirror ratios for ECR sources with a value of 4.3 
over the length of the plasma chamber. The hexapole is cooled at the 
front pieces and on the inner radius by a forced flow of demineralised 
water. The plasma chamber is formed by a 0.4 mm thick stainless steel 
tube of inner diameter of 85 mm and a length of 265 mm axially limited 
by copper inserts. At the position of the hexapole gap the stainless 
steel tube is perforated by a series of  2.5 mm diameter holes allowing 
for radial pumping in addition to axial pumping.
The microwave high frequency power 
 is introduced directly to the plasma chamber with waveguides exhibiting 
 a small angle to the axis. In this way the path of the X-rays was at no point 
 cut by any obstacle which could distort the shape of the response function.

 An extraction voltage of 2 kV had been routinely applied at the  side
 opposite to the crystal spectrometer. The total 
ion current was  measured  as a control  for a stable operation.

 A reference pressure (without plasma) of 1.7 $10^{-7}$ mbar was achieved. Gas 
filling was supplied radially by UHV precision leak valves  through 
the gaps in  the open structure hexapole.  The gas composition was routinely surveyed with a quadrupole 
mass spectrometer.
For an optimised plasma source a drastic increase of the number
 of energetic electrons was discovered which required 
the use of a cleaning magnet installed at a distance of one meter in 
front of the crystal.
\item
 A silicon(111) as well as  a quartz(10$\bar{1}$) crystal were  investigated,
 which had been recently applied 
for measuring  pionic hydrogen transitions  
\cite{stori02}.
   The Bragg angles $\Theta_B$, corresponding to  the M1 transition of helium-like argon
 with an energy of 3.104 keV, are $\Theta_B$= 36.68$^\circ$ for 
the quartz and $\Theta_B$=39.57$^\circ$ 
for the silicon crystal.
 The  crystals  were installed at a distance
 of 2330 mm from the centre of the ECRIT resulting in a position of the 
plasma about 500 mm outside the Rowland circle.

\item
 A Charged Coupled Device (CCD)  pixel detector with a pixel 
size of 40 $\mu$m  x 40 $\mu$m and an energy 
resolution of 140 eV at 3 keV was used to detect the  
X-rays \cite{nelms}. The detector consisted out of six chips with 600 x 600
pixels each resulting in a total height 
of 72 mm and a width of 48 mm. 
 The distance of the CCD detector from the crystal could be changed remotely
 over a length of 86 mm without breaking the vacuum.

 The CCD  detector  
and the associated electronics were  protected against light as well as 
the high frequency stray field  by a 30 $\mu$m thick beryllium window
 installed in the 
vacuum tube in front of the CCD cryostat.

\end {itemize}
\section{Tuning the ECRIT}

\begin{figure}[t]
\centering
\includegraphics[scale=0.7]{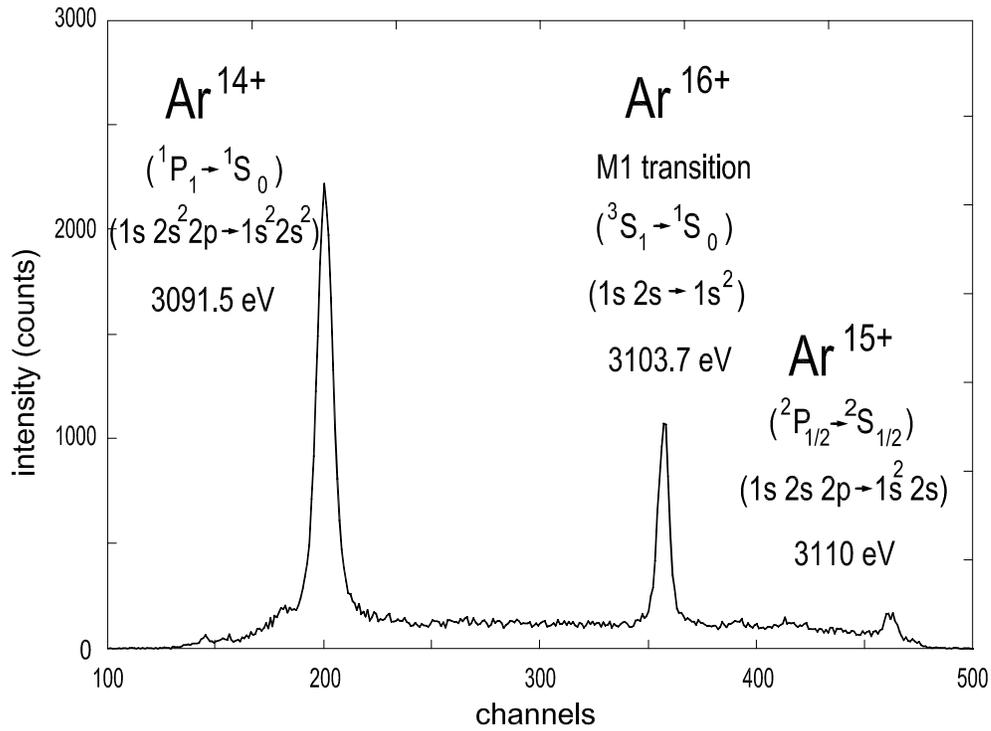}
\caption[]{
The spectral region for argon X--rays from Ar$^{14+}$, Ar$^{15+}$ and Ar$^{16+}$ (M1 transition)
 as emitted from the PSI ECRIT  and measured with
 a Si (111) crystal. The energy values are from \cite{paul1}.
One channel (pixel) corresponds to 79\,meV.  
}
\label{fig:caen4}
\end{figure}

In a first step with low high frequency power and minimum field values around 1.4 
kG the proper working of the whole set-up was checked with X-rays of 
low ionisation degree. In consequence the high frequency power was gradually 
increased and with the Bragg angles set to the proper values for a 
simultaneous observation of the well separable X-rays from both Ar$^{10+}$ 
and Ar$^{11+}$  an optimum for the high frequency power was found at about 
450 Watt. The turn-around time from introducing a change of parameters to 
the evaluation of a measurement was about one minute. Going to Bragg 
angles valid for higher ionisation degrees the strength of the mirror 
field as well as the argon pressure were optimised for a maximum 
intensity of X-rays from highly excited ionised atoms. In addition a gas 
mixture Ar/O$_{2}$ with a mixing ratio of 1:9 was found to be necessary to 
achieve this goal \cite{mix}. The working pressure was 1.4 $10^{-6}$ mbar.
 After a total 
tuning time of 10 hours the M1 transition 
$^{3}S_{1}$$\to$$^{1}S_{0}$ at 3104 eV 
in Ar$^{16+}$could be observed as shown in Figure \ref{fig:caen4}. The
  typical illumination time of the CCD chips of 1 min 
before readout resulted in a high probability of double hits per pixel 
even in the M1 peak. In order to reduce 
the intensity and especially in order to improve the peak/background 
ratio an aluminium collimator was inserted at a distance of 185 mm from 
the centre of the plasma into the direction of the crystal.
 It left a free hole with the dimensions 
16 mm(horizontal) x 10 mm(vertical), which was wide enough not 
to influence the measurement of
 the response function.

A 6.4 GHz 
emitter proved to be  sufficient to provide a highly intensive source of 
M1 X-rays from Ar$^{16+}$ ions in the special hybrid magnetic structure used
  with radial pumping together with  the gas mixing technique. After turning the 
system on, it worked in a stable and reproducible way after about one 
hour warm up time. The observation of the X-rays alone is 
sufficient for a successful tuning in a short time thus making the usual 
momentum analysis of extracted ions superfluous. 
There was no measurable effect of the applied  extraction voltage on 
the intensities of the X-rays for values between 0 and 6 kV,
 which is the maximum achievable voltage.

\section{Response function of the  crystal spectrometer}

With  the ECR source a number of 20000 events was reached for the 
narrow M1 transition of helium-like Argon 
 in about 30   minutes time to be compared with a number 
of 5000 counts reached after  40 hours  with X-rays from pionic 
 carbon formed when using methane gas. A  total of  about 10 hours was needed, however, 
to determine the  spectrometer's response function  in sufficient 
detail  including 
 changes of the distance CCD detector-crystal (focal scans) and changing 
 apertures in front of the crystals. 
 The measurements were confronted 
 with theoretical predictions using the following procedure:
\begin{itemize}
\item The X-ray Oriented Program (XOP) program package  \cite{xop} 
was used as a theoretical 
reference for the present measurements. It provided a theoretical 
rocking curve for a flat ideal crystal.

\item  The theoretical rocking curve  was used as input for a Monte Carlo 
X-ray tracking routine.
 With it the Bragg reflection of a spherically bent 
 crystal was calculated for the geometrical conditions of the special 
 set-up parameters like source diameter, detector position and aperture
 at the crystal. 
\item The output of the tracking routine can be considered to be 
the theoretical limit of the response function which under
 ideal circumstances
 would reproduce the  data.  
In order to allow for additional broadening this response function 
is folded with a Gaussian distribution. Its width was determined  with 
 a least square fitting of the data  using the program package  MINUIT 
from the CERN program library.
The statistics of the Monte Carlo response function was chosen high enough not 
to contribute to the error of the fit.
\end{itemize}

Two types of changes in the experimental set-up were investigated 
 with this procedure. At first  spectra were taken for different distances 
of the CCD detector from the  crystals (focal scans).
In a second step with the CCD detector located  at the focal
 position different apertures  were mounted in front of the crystals.

The focal scans served  two purposes: 

\begin{itemize}
\item to verify the optimum  focal position, which 
is given by $R_C sin\Theta_B$. For this position 
 the width of the response function should be  at minimum. 
The minimum was searched for  by moving the CCD in steps 
of 2-3 mm in  direction to and from the  crystal.
\item to measure the line shape variations as a function of the 
crystal-detector distance which  
 served as a check of  the validity of  X-ray tracking calculations. 
\end{itemize}

\begin{figure}[t]
\centering
\includegraphics[scale=0.7]{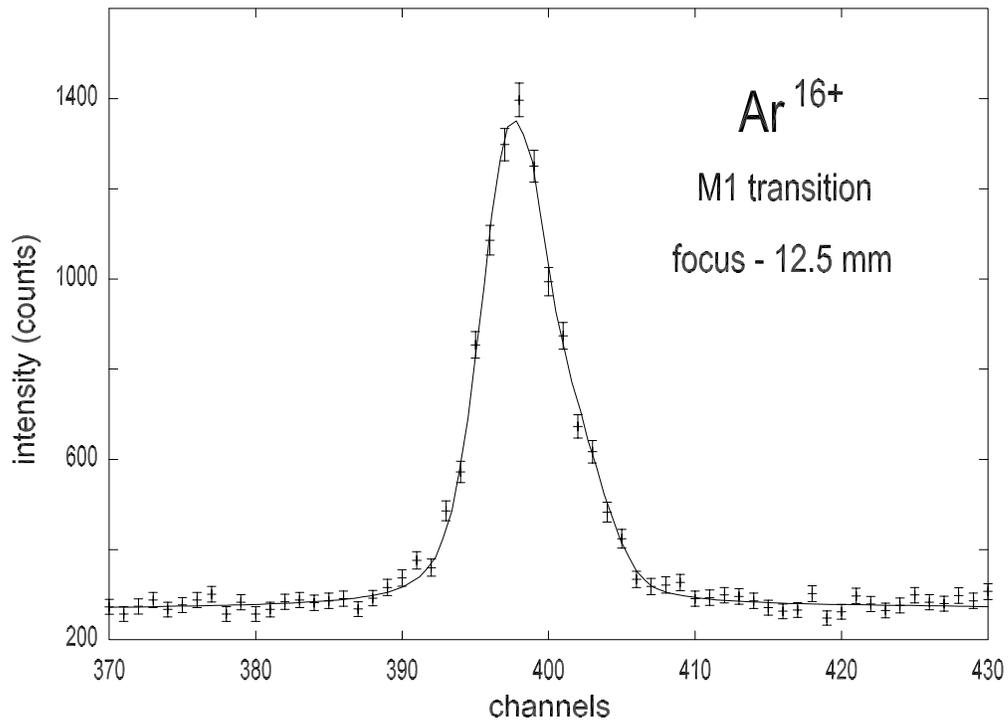}
\caption[]{The spectrum  of the Ar$^{16+}$ 3104 eV M1 line  
measured with a quartz (10$\bar{1}$) crystal is shown together with a fit 
for a detector position of 12.5 mm out of focus in direction of the 
crystal.  A circular  aperture with a diameter of 60 mm 
had been applied. One channel (pixel) corresponds to 93\,meV.  
}
\label{fig:qu10-1_12.5}
\end{figure}

\begin{figure}[t]
\centering
\includegraphics[scale=0.7]{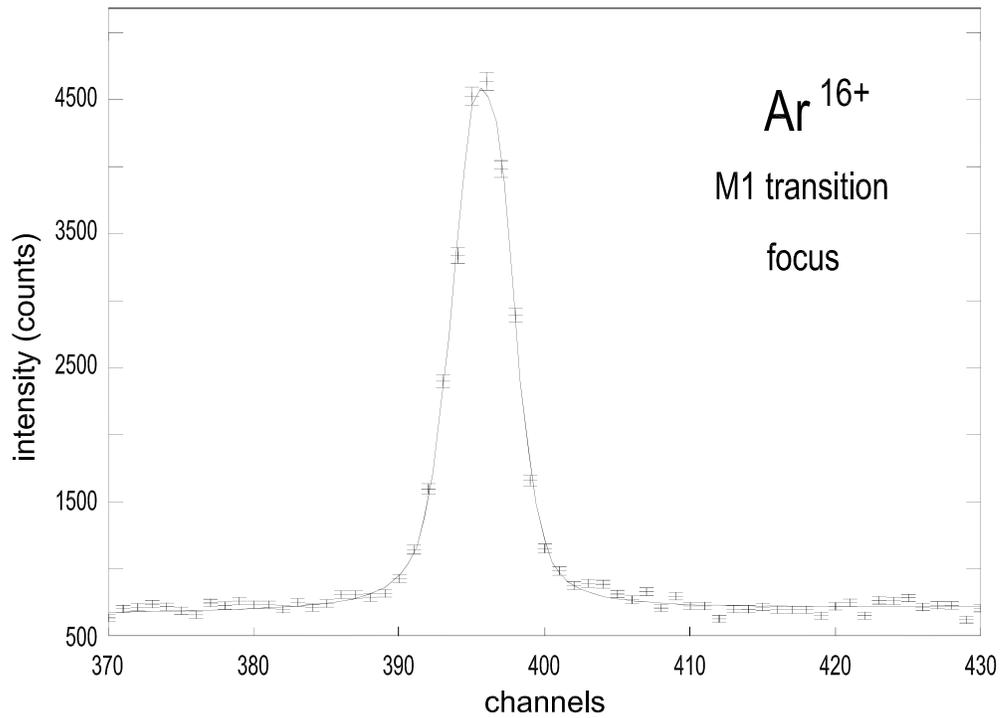}
\caption[]{The spectrum  of the Ar$^{16+}$ 3104 eV M1 line  
measured with a quartz (10$\bar{1}$) crystal is 
shown for the focal position together with a fit.
 A circular  aperture with a diameter of 60 mm 
had been applied. One channel (pixel) corresponds to 93 \,meV.  
}
\label{fig:quc60}
\end{figure}

An example of a spectrum far off focus is shown in
 Figure \ref{fig:qu10-1_12.5} for the quartz  crystal
 with the CCD detector shifted by 12.5 mm from the
focal position into the  direction of the crystal. The FWHM 
of the transition is 657(7)~meV (0.817(9)~m\AA) to be compared 
with the distribution at the focal position with a FWHM value 
of 459(6)~meV (0.591(8)~m\AA) shown in Figure  \ref{fig:quc60}.
The fit allowing for an additional Gaussian broadening 
 reproduces the measured spectrum  in detail.

 In a second step the measured response function is compared with the 
simulated response function, as function of the crystal's reflection 
surface. This is achieved by applying  apertures in front of the crystal
and allows the study of possible irregularities in the response function
due to crystal material imperfections and/or deviations from the spherical
 shape.
Four different apertures were used: two circular ones with openings of diameter
 40 mm and 60 mm, respectively and two rectangular ones 
with horizontal openings 
of 40 mm and 60 mm, respectively, and a free diameter of 95 mm elsewhere. 
 In table \ref{table:1} 
 the results for the theoretical response function are 
confronted with the measurement. Both lineshapes were fitted with a single 
Gaussian and result in the values for the FWHM depicted in columns 2 and 3. 
The  rocking curve width for a plane crystal, as obtained by XOP, fitted by a single 
Gaussian resulted in  FWHM values of 106.6 $\mu$rad for silicon and 
97.4 $\mu$rad for quartz.

 The difference between the  FWHM of the spectrometer's response 
function and the crystal's rocking curve  reflects expected geometrical
 contributions (mainly due to the horizontal extension of the 
crystal-Johann broadening). 
 In order to cope with the difference of the measured response function to the
 theoretical one,  a least square fitting of the data was done with 
the width of 
a Gaussian  folded with the theoretical response function as fit parameter.
 The result of this fit is shown in the fourth column and is considered to be 
the main result of the present investigation.

The Si crystal shows an additional Gaussian contribution  
 of  34.1(1.8) $\mu$rad  as an average of all  apertures chosen.
 This can be interpreted as an overall deviation of 
 the crystal geometry from spherical shape eventually caused  by the 
production process or by  
 a failure in  the crystal material.
 For quartz the three 
smaller apertures show an additional broadening on average of about 24.2(1.8) 
 $\mu$rad, but opening the aperture to 60 mm rectangular 
  results in a considerable worsening. 
This  points to an irregularity in the crystal material or its mounting
 in the upper and/or lower part of the reflecting surface
 and will be investigated in a future measurement  for this special crystal.

The different dispersions
 of  3.756 meV/$\mu$rad for Si and 
 4.166 meV/$\mu$rad for quartz  result in an  
energy resolution  of about  478(6) meV for the Si crystal 
 and 506(8) meV for the quartz crystal, both equipped with 
a 60 mm horizontal aperture. Going to 60 mm circular aperture 
these values reduce to  456(8) meV in the case of Si and to 
459(6) meV in the 
 case of quartz.

The result of a  measurement for a circular aperture with a free diameter 
 of  60 mm is shown in Figure  \ref{fig:quc60} for the quartz crystal
 together with the corresponding fit.

In terms of energy resolution the values for the additional Gaussian 
correspond to   100(8)  meV for quartz  
 and  to 128(6) meV for Si.  Being different they cannot be caused by 
 a common Doppler broadening caused by the motion of the argon ions. 
Moreover it should be recalled that from optical measurements an upper limit 
 for the  Doppler broadening of about 40 meV had been
 obtained \cite{bern,sadeghi}. 

 Compared to the total  
width of the response functions  the   Doppler broadening 
 can be neglected and the present accuracy is sufficient
 for the determination of the strong interaction width in  pionic 
hydrogen.
As the response function has been measured at 3.104 keV  only,  its 
 shape has to be extrapolated to the lower energies of the different 
 pionic hydrogen transitions. If one interpretes the additional Gaussian
 broadening as resulting  from a  geometrical deformation only 
 this extrapolation  is straightforward and the Gaussian  broadening
 can be folded into the rocking curve before the X-ray tracking is performed.

\section{Conclusion and Outlook}

The characterisation of two large area spherically bent crystals
 was performed with
 M1 X-rays from helium-like argon with  two 
 orders of magnitude higher statistics than achievable 
 with exotic atoms. The geometrical conditions of the characterisation 
were equivalent to  the pionic hydrogen experiment. 
The measurement could be understood in terms of a Monte-Carlo X-ray 
tracking routine  allowing, however, for an additional Gaussian broadening. 
>From this a prescription for extrapolating the result to different energies 
became possible.

In the future it is planned to improve on the peak/background ratio by 
insertion of a proper slit of tapered material with a hole of 1 mm(vertical) x 
30 mm (horizontal) near the plasma chamber in the direction of the 
crystal.

In addition the achievable pressure will be optimised by reducing the 
surface of the iron insertion pieces drastically. 
The use of  M1 transitions from    
helium-like  chlorine ( supplied from the gaseous compound 
CHClF$_2$, chlorodi\-fluoromethane )  and
 helium-like sulfur from SO$_2$ is
 envisaged for future experiments as a countercheck of the 
assumptions and conclusions made here. The 
  energies at 2.430 and 2.756 keV, respectively, are close to 
 the 2p-1s and the 3p-1s transition energies in pionic hydrogen.

Additionally a thorough investigation of X-ray intensities as a 
function of its radial origin inside the plasma  is feasible 
with the special set-up of the crystal spectrometer.
The described set-up allows to  
determine the Doppler broadening from the ion velocity 
distribution directly  
 with crystals with better resolution. Similarly 
 the observation of such a  broadening as a function of the power
 and frequency injected into the  ECR source  may be studied.

\section{Acknowledgements}

The suggestions and the help of D. Hitz and K. Stiebing in the 
preparatory phase of the experiment are warmly acknowledged. We also 
thank H. Reist for offering  the 6.4 GHz emitter for free.
Special thanks go to the Carl Zeiss Company in Oberkochen, Germany, which 
manufactured the Bragg crystals.
Laboratoire Kastler Brossel is Unit\'e Mixte de Recherche du CNRS 
n$^{\circ}$ 8552.

\newpage
\section*{Figures}

Fig. 1. Set up of the PSI ECRIT together with the Bragg crystal spectrometer

Fig. 2. The spectral region for argon X--rays from Ar$^{14+}$, Ar$^{15+}$ and Ar$^{16+}$ (M1 transition)
 as emitted from the PSI ECRIT  and measured with
 a Si (111) crystal. The energy values are from \cite{paul1}.
One channel (pixel) corresponds to 79\,meV.

Fig. 3. The spectrum  of the Ar$^{16+}$ 3104 eV M1 line  
measured with a quartz (10$\bar{1}$) crystal is shown together with a fit 
for a detector position of 12.5 mm out of focus in direction of the 
crystal.  A circular  aperture with a diameter of 60 mm 
had been applied. One channel (pixel) corresponds to 93\,meV.

Fig. 4. The spectrum  of the Ar$^{16+}$ 3104 eV M1 line  
measured with a quartz (10$\bar{1}$) crystal is 
shown for the focal position together with a fit.
 A circular  aperture with a diameter of 60 mm 
had been applied. One channel (pixel) corresponds to 93 \,meV.

\newpage
\begin{table}[t]
\caption{ 
 The FWHM in $\mu$rad of the theoretical response function 
  is compared with the measurement  for a Si (111) as well as 
a quartz (10$\bar{1}$) crystal
 for different openings (rectangular or circular) of the diffraction surface.
 The last column shows 
the  FWHM for  a Gaussian to be folded
 with the theoretical response 
function in order to fit the measurement best.
\vspace{1cm}
}                       
 
\label{table:1}
\newcommand{\m}{\hphantom{$-$}}
\newcommand{\cc}[1]{\multicolumn{1}{c}{#1}}
\renewcommand{\tabcolsep}{1pc} 
\renewcommand{\arraystretch}{1.2} 
\begin{tabular}{@{}lllll}
\hline  
Crystal/aperture         & \m Theor. resp. fct. & \m Measurement & \m  Gaussian\\
\hline
Si/Rect. 60        & \m 116.0(0.5)   & \m 127.4(1.7)   & \m  38.0(4.0) \\
Si/Circ. 60           & \m 114.5(0.5)   & \m 121.5(2.2)   & \m   30.3(3.5) \\
Si/Rect. 40        & \m 108.5(0.6)   & \m 119.8(2.2)   & \m  34.6(3.2) \\
Si/Circ. 40           & \m 108.7(0.6)   & \m 117.5(2.1)   & \m  34.2(2.7) \\
\hline
quartz/Rect. 60    & \m 109.4(0.3)       & \m 121.5(1.8)  & \m 38.0 (4.1) \\
quartz/Circ. 60       & \m 106.2(0.3)       & \m 110.2(1.4) & \m 21.5 (3.2)  \\
quartz/Rect. 40    & \m 98.8(0.4)        & \m 104.2(1.7) & \m 25.1 (2.9)  \\
quartz/Circ. 40       & \m 98.4(0.4)        & \m 104.2(1.6) & \m 25.6 (3.1)  \\
\hline
\end{tabular}\\[2pt]
\end{table}

\newpage

\pagebreak
\newpage

\pagebreak

\newpage

\end{document}